# Moiré-enabled optical vortex with tunable topological charge in twisted bilayer photonic crystals

Tiancheng Zhang[1,7], Li Lei[1,7], Changhao Ding[1,7], Fanhao Meng[2], Qicheng Jiang[2], Lijie Li[1], Scott Dhuey[3], Jingze Yuan[2], Zhengyan Cai[2], Yi Li[2], Jingang Li[4], Costas P. Grigoropoulos[4], Haoning Tang[5], Jie Yao[1,2,6,*]

[1]Applied Science and Technology Graduate Group, University of California, Berkeley, CA, 94720, USA.

[2]Department of Materials Science and Engineering, University of California, Berkeley, CA, 94720, USA.

[3]The Molecular Foundry, Lawrence Berkeley National Laboratory, Berkeley, CA, 94720, USA.

[4]Department of Mechanical Engineering, University of California, Berkeley, CA, 94720, USA.

[5]Department of Electrical Engineering and Computer Sciences, University of California, Berkeley, CA, 94720, USA.

[6]Materials Sciences Division, Lawrence Berkeley National Laboratory, Berkeley, CA, 94720, USA.

[7]These authors contributed equally: Tiancheng Zhang, Li Lei, Changhao Ding

*e-mail: yaojie@berkeley.edu

**Abstract**

The orbital angular momentum (OAM) of light is a versatile degree of freedom with transformative impact across optical communication, imaging, and micromanipulation. These applications have motivated a growing demand for compact, reconfigurable vortex arrays with tunable topological charge, yet integrating these functionalities into nanophotonic platforms remains elusive. Among possible strategies to meet this challenge is exploiting the twist degree of freedom in layered structures, which enables both emerging moiré physics and unprecedented reconfigurability of photonic and electronic properties. Here, we harness these capabilities in twisted bilayer moiré photonic crystals (TBMPCs) to realize vortex array generation with tunable OAM, demonstrated both analytically and experimentally. Central to this advancement is a new class of quasi-bound state in the continuum: Bessel-type modes emerging from moiré-induced interlayer coupling, which generate vortex beams with tailored spiral phase distributions. We experimentally demonstrate vortex beams spanning eight OAM orders, from −3 to 4, and achieve selective excitation of distinct topological charges at a fixed telecommunication wavelength by tuning the interlayer separation and twist angle. Furthermore, localized Bessel-type modes at AA stacking regions can be excited nonlocally across the moiré superlattice, enabling vortex array generation. Our work offers new insights into moiré physics and introduces an innovative approach for future multiplexing technology integrating OAM, wavelength, and spatial division.

**Introduction**

Angular momentum—a fundamental physical quantity governing rotational phenomena—pervades scales from spiral galaxies to quantum vortices. In photonics, the orbital angular momentum (OAM) of light and associated optical vortices have attracted significant research interest. As a degree of freedom independent of wavelength and polarization, OAM has revolutionized multiple fields, including optical communication[1-3], super-resolution microscopy[4,5], quantum information processing[6,7], micromanipulation[8-10], and metrology[11-13]. In optical communications, distinct OAM states in vortex beams provide independent transmission channels for bandwidth enhancement[1-3], and can also serve as the signal-encoding degree of freedom[14]. For micromanipulation, different OAM states substantially alter light-matter interactions[15]. These OAM-based applications underscore the critical need for vortex sources with tunable topological charge (hereafter referred to as OAM-tunable)[16]. While solutions based on large-scale systems exist[17], the demand for enhanced compactness and spatial density—crucial for higher channel capacity and functionality—drives the pursuit toward OAM-tunable vortex arrays in nanophotonic systems. Current micro/nanoscale devices, however, cannot simultaneously deliver OAM tunability and vortex array generation[18-23]. As a result, new concepts and approaches are being sought to realize reconfigurable vortex arrays with tunable OAM.

The unique advantages of moiré photonics offer new opportunities. Inspired by moiré van der Waals materials, which have led to the observation of fractional Chern insulators[24], moiré excitons[25], superconductivity[26], and polar vortices[27,28], twisted bilayer moiré photonic crystals

(TBMPCs) have also drawn significant attention from the optical community. Multiple emerging phenomena have been demonstrated in the TBMPC system, including moiré flat bands[29-31], energy localization[32,33], moiré nonlinearity[34], moiré nanolaser arrays[35], and optical solitons[36]. Single-order optical vortex generation has also been theoretically predicted[37]. Moreover, by integrating with micro-electromechanical systems (MEMS), TBMPCs function as highly reconfigurable optical platforms with dynamically tunable configurations[38], demonstrating strong potential for wide tunability in future moiré photonics applications.

In this work, we integrate the three key merits of TBMPCs—reconfigurable operation, moiré superlattice engineering, and intrinsic vortex generation—to theoretically and experimentally demonstrate the generation of OAM-tunable vortex arrays. Although vortex beam generation has been theoretically predicted in TBMPCs[37], the OAM order of the radiated beam is limited and cannot be tuned. Here, we reveal a new class of quasi-bound state in the continuum (quasi-BIC) in TBMPCs (termed Bessel-type modes) in each moiré supercell, which generates vortex beams spanning a wide range of OAM orders. These quasi-BICs emerge through unique interlayer coupling, exhibiting multi-order dynamic phase vortices centered at the AA region of each moiré supercell. When combined with the Pancharatnam-Berry (PB) phase from spatially varying polarization, they generate out-of-plane vortex beams with distinct OAM. Experimentally, we demonstrate vortex radiation spanning eight OAM orders from −3 to 4. By modulating the interlayer separation and twist angle, we achieve single-wavelength OAM switching—a capability unattainable in most conventional platforms. Furthermore, vortex generation from AA regions across the moiré superlattice forms a vortex array, while also allowing selective single-beam vortex radiation. The system operates without stringent constraints on incident beam position or tilt angle, significantly relaxing alignment tolerances. Collectively, TBMPCs establish a robust platform for high-spatial-density vortex generation featuring exceptional tunability.

**Results**

**Methodology and vortex array generation**

Vortex beam generation fundamentally emerges from two distinct phase contributions: the dynamic phase, representing the phase evolution during propagation[39], and the PB phase, arising from spatially varying polarization orientations[40]. The independent control over these components offers unprecedented flexibility for phase modulation in vortex beams[16]. In TBMPCs, constructed by slightly twisted photonic crystal (PhC) slabs (Fig. 1a), theoretical analysis has demonstrated that a twist-enabled quasi-bound state in the continuum (quasi-BIC) in the bilayer structure can radiate vortex beams. The OAM order $q$ of the radiation was limited to $q = \pm m$, where $m$ is the topological charge of the at-$\Gamma$ bound state in the continuum (BIC) of the single layer PhC[34,37]. This effect relies exclusively on the PB phase which is difficult to tune. In contrast, the Bessel-type modes in this work—identified as truncated Bessel functions in reciprocal space—introduce dynamic phase vortices, enabling OAM modulation through the combined use of dynamic and PB phase. Bessel-type modes can be excited remotely: under illumination such as a Gaussian beam, delocalized guided resonances activate localized Bessel-type modes (Fig. 1b), which in turn radiate vortex beams from the AA regions of the moiré superlattice. Together, the dynamic-phase-induced OAM modulation and nonlocal excitation of Bessel-type modes enable vortex array generation with OAM orders extending beyond the constraint $q = \pm m$ (Fig. 1a).

Prior to the analysis of Bessel-type modes, we present direct experimental evidence of the nonlocal excitation mechanism: vortex array generation from the TBMPC (Fig. 1c). A schematic of the optical setup is shown in Extended Data Fig. 1. A TBMPC structure is prepared with one PhC layer extending beyond the overlapped TBMPC region (Extended Data Fig. 9). As indicated by the blue dashed circle in Fig. 1c, the incident beam illuminates the single-layer PhC region. Nevertheless, multiple donut-shaped radiation spots emerge at distinct AA regions within the TBMPC area. To confirm their vortex nature and directly measure OAM, we measure the far-field beam profile and coaxial interference pattern of a single vortex (Figs. 1d and 1e). These results clearly reveal the characteristic donut-shaped intensity distribution and spiral interference structure—definitive signatures of vortex beams. The spatial separation between the incident beam and the vortex array simultaneously validates the nonlocal excitation mechanism in Fig. 1b and confirms that vortex generation originates solely from the TBMPC rather than the single-layer structure. The observation of vortex radiation with first-order OAM—evidenced by the single spiral arm in the interference pattern—further reinforces this conclusion, as such a state is fundamentally inaccessible in single-layer PhCs[20].

**Bessel-type modes in TBMPCs**

In this section, we theoretically investigate the formation of Bessel-type modes and their radiation properties, showing that these modes introduce an additional dynamic phase to the radiated field, enabling access to distinct OAM orders. Prior studies on moiré structures have primarily focused on the hybridized guided resonances near the K-point, where Dirac cones are located, while leaving hybridization around the Γ-point relatively unexplored. Here, we establish a model for the non-trivial coupling mechanism in TBMPCs near the Γ-point that describes the formation of Bessel-type modes.

Coupled-mode theory[41,42] is used as it has been proven to effectively describe the photonic behavior of TBMPCs[29,34,37]. In a monolayer PhC, the amplitude of the guided resonance with wavevector $\boldsymbol{k}$ and eigenfrequency $\omega(\boldsymbol{k})$ (hereafter denoted as mode $\boldsymbol{k}$) is described by $a(\boldsymbol{k})$. When two PhCs (PhC1 and PhC2) are stacked with subwavelength separation, their guided resonances—mode $\boldsymbol{k_1}$ in PhC1 and mode $\boldsymbol{k_2}$ in PhC2—interact with a coupling coefficient $\zeta(\boldsymbol{k_1}, \boldsymbol{k_2})$. For non-twisted bilayer PhCs, momentum conservation requires $\zeta(\boldsymbol{k_1}, \boldsymbol{k_2}) \neq 0$ only when $\boldsymbol{k_1} = \boldsymbol{k_2}$ within the first Brillouin zone[37]. The hybridization of two guided resonances with identical wavevector yields a trivial bilayer guided resonance (Fig. 2a).

The scenario diverges significantly in TBMPCs, where the coupling strength remains finite for $\boldsymbol{k_1} \neq \boldsymbol{k_2}$ and satisfies $\zeta(\boldsymbol{k_1}, \boldsymbol{k_2}) = \zeta(\boldsymbol{k_1} - \boldsymbol{k_2})$ near the Γ-point. This allows the hybridization of guided resonances with different wavevectors, forming localized eigenmodes

(Figs. 2b and 2c). The hybridized mode near the Γ-point can be obtained by solving the coupled-mode equations:

$$\begin{cases} \dfrac{da_1(\boldsymbol{k}_1)}{dt} = -i\omega_\Gamma a_1(\boldsymbol{k}_1) + i \iint \zeta(\boldsymbol{k}_1 - \boldsymbol{k}_2) a_2(\boldsymbol{k}_2) d\boldsymbol{k}_2 \\ \dfrac{da_2(\boldsymbol{k}_2)}{dt} = -i\omega_\Gamma a_2(\boldsymbol{k}_2) + i \iint \zeta(\boldsymbol{k}_2 - \boldsymbol{k}_1) a_1(\boldsymbol{k}_1) d\boldsymbol{k}_1 \end{cases} \quad (1)$$

where $\omega_\Gamma$ denotes the eigenfrequency of the single-layer at-Γ BIC. This equation is derived under the condition $|\boldsymbol{k}|L \ll 2\pi$, where $L$ is the lattice constant of the single-layer PhC. Observing that the subscripts "1" and "2" are exchangeable, we set $a_2(\boldsymbol{k}) = -a_1(\boldsymbol{k}) = -a(\boldsymbol{k})$ (see Methods), simplifying equation (1) to a standard integral equation:

$$\Delta\omega a(\boldsymbol{k}) = \iint \zeta(\boldsymbol{k} - \boldsymbol{k}') a(\boldsymbol{k}') d\boldsymbol{k}' \quad (2)$$

where $\Delta\omega = \omega_B - \omega_\Gamma$ (with $\omega_B$ being the eigenfrequency of the new solutions) denotes the eigenfrequency difference. The solution of equation (2) manifests as a class of Bessel functions in reciprocal space: $a(\boldsymbol{k}) = j_{q_d}(b_n|\boldsymbol{k}|)e^{iq_d\phi_k}$. Here $b_n$ is the radial scaling factor, $q_d$ is an integer azimuthal order, and $\phi_k$ is the azimuthal coordinate in reciprocal space, respectively. Since equation (1) assumes small $|\boldsymbol{k}|$ near the Γ-point, it fails for large $|\boldsymbol{k}|$. The solutions therefore become truncated Bessel functions, parameterized by radial ($n$) and azimuthal ($q_d$) orders[43]. We refer to these eigenmodes as Bessel-type modes. Their corresponding real-space field distributions also exhibit vortex profiles centered at the AA region of each moiré supercell (Extended Data Fig. 7). As the twist angle decreases, the Bessel-type modes become less localized, eventually transforming into the at-Γ BIC at zero twist angle. This evolution identifies them as a new class of quasi-BICs.

To examine our theoretical framework, we performed simulations of the eigenmodes within the TBMPC system near the Γ-point. The top (bottom) row in Fig. 2d shows the simulated complex mode amplitude in the reciprocal space for $q_d = 1$ ($q_d = 0$) and $n = 1, 2, 3$ modes. Unlike the $q_d = 0$ modes, which exhibit a trivial dynamic phase, the $q_d = 1$ modes feature phase singularities at the Γ-point. Remarkably, these simulation results are in excellent agreement with the analytically derived truncated Bessel functions, as confirmed by the side-by-side comparison in Fig. 2e. For quantitative validation, Fig. 2f plots $a(\bm{k})$ along the Γ-M direction for both simulated and theoretical results. The two curves in each plot match closely for small $|\bm{k}|$ and show finite but small deviation for $|\bm{k}| > 0.25\pi/L$, consistent with the breakdown of the small-$|\bm{k}|$ approximation. These simulation results not only validate the theoretical analysis but also directly confirm the existence of Bessel-type modes in the TBMPC system.

We proceed to analytically derive the far-field radiation from these Bessel-type modes, shaped by both their dynamic phase and the PB phase arising from spatially varying states of polarization (SOPs). Since the single-layer PhC is protected by $C_2^z T$ symmetry, its far-field SOPs are linearly polarized[44] (Extended Data Fig. 6), oriented at an angle $\theta(\bm{k})$ relative to the x-axis. Around the at-Γ BIC, $\theta(\bm{k}) = m\phi_k + \theta_0$, where $m$ denotes the topological charge of the BIC, and $\theta_0$ is a constant[44]. The explicit expression of the radiated field can be expressed by the temporal coupled-mode theory (TCMT) equation[37] : $|E_{\text{out}}(\bm{k})\rangle = D(\bm{k})a(\bm{k}) \cdot \begin{pmatrix} \cos(\theta(\bm{k})) \\ \sin(\theta(\bm{k})) \end{pmatrix}$, where $|E_{\text{out}}(\bm{k})\rangle$ is the Jones vector of the radiated field, $D(\bm{k})$ is the radiation coefficient with a trivial phase distribution[20], and $a(\boldsymbol{k})$ is the amplitude of the Bessel-type mode in reciprocal space. Decomposing the Jones vector $\begin{pmatrix} \cos(\theta(\boldsymbol{k})) \\ \sin(\theta(\boldsymbol{k})) \end{pmatrix}$ into left- and right-circularly polarized (LCP/RCP) unit vectors ($|L\rangle$/$|R\rangle$) yields:

$$|E_{\text{out}}(\boldsymbol{k})\rangle = D(\boldsymbol{k})a'(|\boldsymbol{k}|) \cdot \left( e^{i(q_d-m)\phi_k - i\theta_0}|L\rangle + e^{i(q_d+m)\phi_k + i\theta_0}|R\rangle \right) \quad (3)$$

where $a'(|\boldsymbol{k}|)$ represents the Bessel-type mode amplitude along the radial axis. These results indicate that radiation from the Bessel-type modes comprises vortex beams with OAM orders $q = q_d - m$ (LCP) and $q = q_d + m$ (RCP). Defining the PB phase order as $q_{\text{PB}} = -m$ (LCP) and $q_{\text{PB}} = m$ (RCP), the OAM order follows: $q = q_d + q_{\text{PB}}$. Figures 2g and 2h schematically illustrate this principle: while radiation from Bessel-type mode with $q_d = 0$ yields vortex beam with OAM order 1, radiation from $q_d = 1$ mode is expected to exhibit OAM order 2, which is verified by simulation (Extended Data Fig. 7) and experiment (see next section).

Bessel-type modes are spectrally distinguishable by their eigenfrequencies. Under Gaussian beam illumination, the mode intensity spectra (Extended Data Fig. 4) reveal the sequential emergence of distinct $q_d$ Bessel-type modes and a monotonic redshift of peak wavelengths with increasing $|q_d|$. All observed modes in the spectra have radial order $n = 0$, as modes with $n > 0$ exhibit negligible excitation efficiency under Gaussian beam illumination; therefore, subsequent analysis focuses exclusively on $n = 0$ modes. By tuning the interlayer separation or twist angle of the TBMPC, the mode intensity spectrum can be shifted to align different $q_d$ modes at a target wavelength $f_0$ (Fig. 2i). Switching between these modes thus enables dynamic control of OAM at a fixed wavelength.

**Sample fabrication and experimental realization**

To experimentally realize the proposed phenomenon, we design and fabricate two identical single-layer PhC slabs operating at telecommunication wavelengths and stack them with a slight twist to form the TBMPC. Figure 3a displays a scanning electron microscope (SEM) image of the fabricated single-layer PhC slab, consisting of a 223-nm-thick silicon layer periodically etched on a quartz substrate. Circular etched holes (302 nm in diameter) are arranged in a honeycomb lattice with a 665-nm lattice constant. The two PhC slabs are bonded through a flip-chip bonding technique, which enables high-precision control of the twist angle[45], with a 280-nm-thick PMMA spacer separating the two layers. A moiré pattern appears in the optical microscope image after bonding at a twist angle $\theta = 1.7°$ (Fig. 3b). Angle-resolved transmittance spectra of the fabricated TBMPC are shown in Fig. 3c, with a magnified view in Fig. 3d resolving two bands within the spectral range, both corresponding to a BIC order $m = 1$. We selectively utilize the upper band to realize vortex beam radiation.

The imaging setup is shown in Extended Data Fig. 1, with a detailed explanation in the Methods section. The system first operates in imaging mode with no reference beam applied. We use the linear polarizer (LP)1, quarter-wave plate (QWP)1, and objective lens (OL)1 to create a convergent RCP incident beam that excites the Bessel-type modes. The resulting radiation of the modes and directly transmitted light are collected by OL2. However, only the

LCP component of the radiation reaches the infrared (IR) camera, as QWP2 and LP2 block both the RCP radiation component and the transmitted light. This configuration ensures that the camera receives enhanced signals when Bessel-type modes are excited, making the cross-polarization transmission directly correspond to the Bessel-type mode intensity. By scanning wavelengths from 1,519 nm to 1,530 nm, we obtain the cross-polarization transmission spectrum, which reveals three resonant peaks at 1,520.7 nm, 1,525.2 nm, and 1,529.1 nm (Fig. 3e, labeled I, II, and III), with quality factors of $2.6 \times 10^3$, $1.9 \times 10^3$, and $1.3 \times 10^3$, respectively. To verify vortex beam generation, we then measure the corresponding far-field radiation. Figure 3f (top panels) shows three donut-shaped beam profiles corresponding to the resonant peaks in Fig. 3e. To directly confirm the vortex nature and measure the OAM order, we introduce a reference beam into the imaging system, switching to interferometric mode. The resulting interference patterns for peaks I, II, and III (Fig. 3f, middle panels) exhibit one, two, and three spiral arms, confirming vortex beams with OAM order $q = -1, -2,$ and $-3$, respectively[19,20].

To demonstrate the independence of the dynamic phase and the PB phase in determining the radiated OAM order, we switch the excitation to LCP and the reception to RCP. Notably, the peak positions in the cross-polarization transmission spectrum remain unchanged under this polarization reversal. Figure 3g shows the beam profiles from peaks I and II under the exchanged polarization: a vortex beam with $q = 1$ and a Gaussian-like beam without phase singularity ($q = 0$). According to our theory, the OAM order follows: $q = q_\mathrm{d} + q_\mathrm{PB}$. With $q_\mathrm{PB} = -1$ for LCP and $q_\mathrm{PB} = 1$ for RCP, peak I, II, and III in Fig. 3e correspond to Bessel-type modes with $q_\mathrm{d} = 0$, $-1$, and $-2$, respectively. These experimental results show excellent agreement with the simulations (Extended Data Fig. 3). Consistently, simulated radiation phases (Fig. 3f and 3g, bottom panels) match the experimental orders. This triangular consistency—theory, simulation, and experiment—confirms our theoretical framework, verifies the existence of Bessel-type modes, and directly demonstrates multi-OAM vortex beams in TBMPCs.

Bessel-type modes with different $|q_\mathrm{d}|$ are spectrally distinguishable by their eigenfrequencies, yet modes with opposite $q_\mathrm{d}$ remain degenerate. The absence of positive-$q_\mathrm{d}$ modes in previous measurements (Figs. 3e-g) reveals an intriguing unidirectional radiation phenomenon. As shown in Extended Data Fig. 8 and described in detail in the Methods section, the positive-$q_d$ modes primarily radiate in the −z direction, while negative-$q_d$ modes radiate toward +z. This explains the absence of positive-$q_d$ modes in typical +z detection schemes. Consequently, at a fixed wavelength and direction, radiation is dominated by a single Bessel-type mode, eliminating phase distortions caused by multi-OAM vortex beam superposition. This unidirectional radiation is twist-angle controlled: by reversing the twist angle from $\theta$ to $-\theta$, converting the TBMPC system to its enantiomorph, radiation in the +z direction becomes dominated by the positive-$q_\mathrm{d}$ modes. We experimentally verify this by fabricating a TBMPC with $\theta = -1.7°$. Figure 3h shows the cross-polarization transmission spectrum, featuring three peaks at 1,520.4 nm, 1,524.7 nm, and 1,527.5 nm (Fig. 3h, labelled as IV, V, and VI), respectively. They correspond to $q_\mathrm{d} = 0$, 1, and 2 Bessel-type modes, respectively, verified via radiation profiles and interferograms in Figs. 3i and 3j. Furthermore, around the $m = -2$ at-$\Gamma$ BIC, we also observed vortex beams radiation from Bessel-type modes with $q = 2, 3, 4$ (LCP) and $-2, -1, 0$ (RCP) (Extended Data Fig. 2). Together, these results demonstrate vortex beam generation with OAM orders spanning from $-3$ to $4$ in TBMPCs (from $-5$ to $6$ in simulations), substantially extending the OAM range beyond the $q = \pm m$ limitation.

**OAM-tunable vortex beam generation**

We have now experimentally demonstrated the generation of vortex beams carrying OAM orders ranging from $q = -3$ to $4$ within the 1,519 nm to 1,530 nm spectral range. This capability establishes the foundation for constructing OAM-tunable vortex sources. As described by Fig. 2i, adjusting the twist angle $\theta$ and interlayer separation $t$ of the TBMPC shifts the resonance peaks in the cross-polarization transmission spectrum, thereby positioning Bessel-type modes with different $q_d$ at a designated wavelength. Experimental validation of this tuning mechanism is provided in Fig. 4a, which displays the cross-polarization transmission spectra under four different configurations: b: ($\theta = 1.7°, t = 280$ nm), c: ($\theta = -1.8°, t = 340$ nm), d: ($\theta = -1.8°, t = 280$ nm), and e: ($\theta = -1.8°, t = 240$ nm). When operating at a fixed wavelength $\lambda = 1{,}525.4$ nm, these configurations (b, c, d, and e) excite Bessel-type modes with $q_\mathrm{d} = -1, 0, 1$ and $2$, respectively. Applying the equation: $q = q_\mathrm{d} + q_\mathrm{PB}$, we expect to observe RCP vortex beams with OAM $q = 0, 1, 2$ and $3$ among these configurations, which is confirmed by the beam profiles and interferograms in Figs. 4b-e. Simulation reveals a significantly broader tuning range, with nine distinct OAM orders from $q = -5$ to $3$ at a single wavelength (Extended Data Fig. 5). The twist angle control is implemented via flip-chip bonding, while the interlayer separation modulation is achieved by adjusting the PMMA spin-coating speed (governing the polymer thickness), and can also be fine-tuned by the pressure applied to the TBMPC. These results demonstrate OAM tunability at a fixed wavelength through twist-angle and interlayer-separation control, establishing TBMPCs as a robust platform for OAM-tunable vortex beam generation. When combined with the nonlocal excitation mechanism enabling vortex array formation (Fig. 1c), this approach achieves a vortex array with tunable OAM in a compact nanophotonic system.

**Conclusion**

In summary, we demonstrate the tunable topological charge vortex array generation in twisted bilayer moiré photonic crystals (TBMPCs). Theoretically, we reveal a new class of quasi-bound states in the continuum (quasi-BICs) for moiré systems—Bessel-type modes whose mode distributions are truncated Bessel functions in the reciprocal space and possess dynamic phase vortices centered at the AA stacking regions in real space. The tunable dynamic phase orders of the Bessel-type modes enable vortex beam generation with multiple distinct orbital angular momenta (OAM), surpassing the intrinsic orders of bound states in the continuum (BICs). We fabricate devices operating at telecommunication wavelengths (near 1525 nm) and experimentally observe vortex beams spanning OAM orders from –3 to 4. Moreover, we experimentally demonstrate switching between different OAM states by modulating the twist angle and interlayer separation of the TBMPC without changing the operating frequency. In addition, vortex beams originating from AA regions of the moiré superlattice collectively form a vortex array, with the flexibility to selectively excite single-beam vortex when desired. Looking ahead, integrating TBMPCs with micro-electromechanical systems (MEMS) and active materials could enable reconfigurable coherent vortex microlaser arrays with tunable OAM and controllable spatial distribution. This novel nanophotonic approach for generating vortex arrays with tunable OAM will advance structured light control and offer broad potential for applications in optical communication, micromanipulation, and quantum information processing.

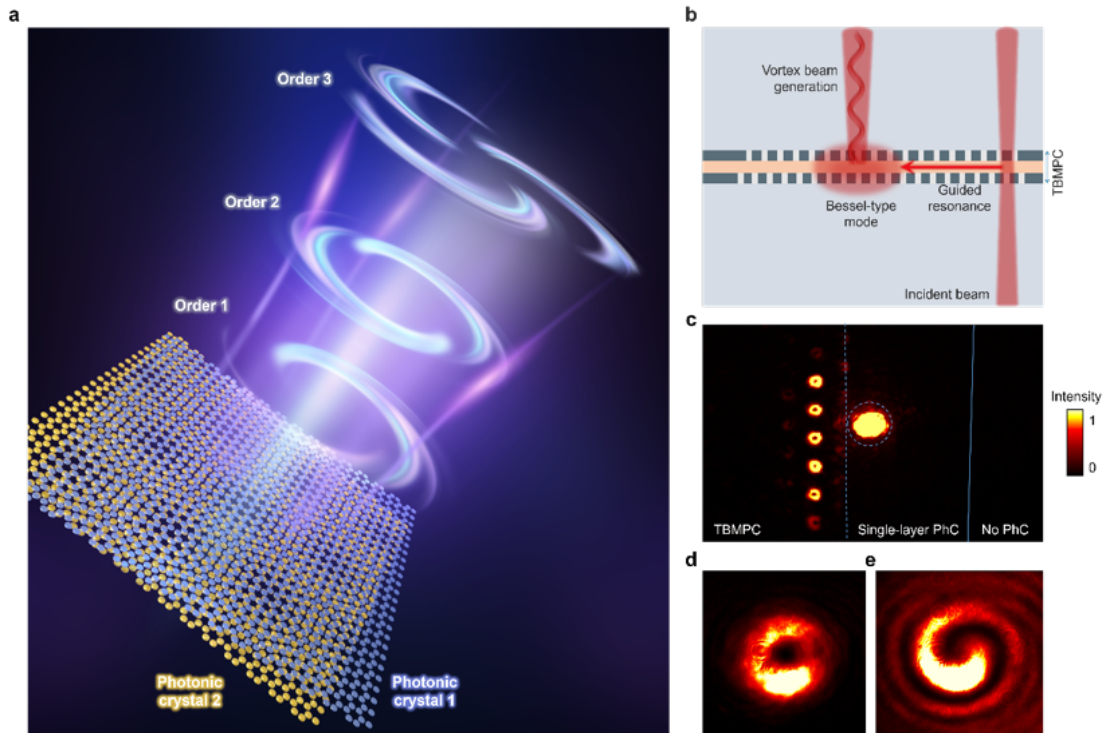

**Fig. 1 | Generation of vortex arrays with tunable topological charge in twisted bilayer moiré photonic crystals (TBMPCs). a,** Schematic of vortex beam generation with tunable topological charges of orbital angular momentum (OAM) from a TBMPC. **b,** Schematic illustrating the nonlocal excitation of Bessel-type modes and the resulting vortex beam radiation from the TBMPC. **c,** Measured intensity distribution of the generated vortex array. The blue dashed line indicates the boundary between the TBMPC and the single-layer photonic crystal (PhC). The blue solid line denotes the boundary of the single-layer PhC. The blue dashed circle denotes the position of the incident beam. **d, e,** Beam profile (**d**) and coaxial interference pattern (**e**) of the right-circularly polarized (RCP) component of a single vortex beam in the far field.

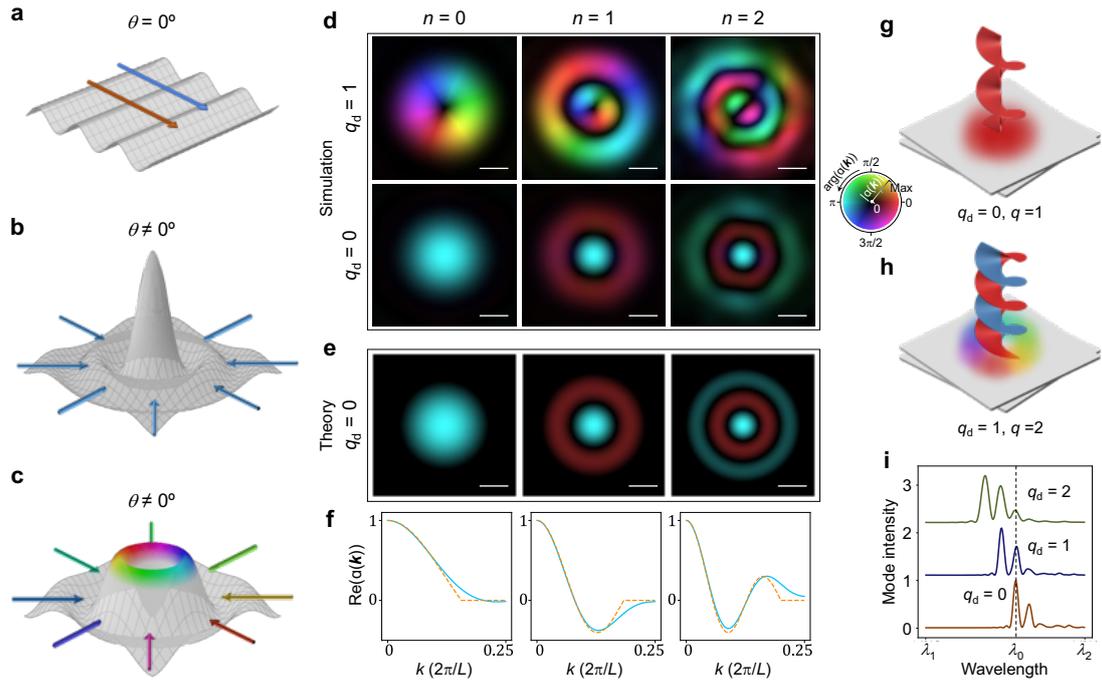

**Fig. 2 | Bessel-type modes in the TBMPC. a,** Schematic of the hybridization of two guided resonances with the same wavevector in a non-twisted bilayer PhC. The arrows indicate the wavevector directions, and their colors correspond to the phases. **b, c,** Schematic of the formation of localized modes with trivial (**b**) and nontrivial (**c**) phase distributions through the hybridization of multiple guided resonances with different wavevectors in TBMPCs. The height of the surfaces represents the intensity distribution of the modes. The arrows denote the wavevector directions, and their colors indicate the corresponding phases. In **c**, the color scale on the surface indicates the phase value at the intensity peaks. **d,** Amplitude and phase of the simulated Bessel-type modes in reciprocal space for $q_d = 1$ (top) and $q_d = 0$ (bottom), with $n = 0, 1,$ and $2$. Scale bar: $0.2\pi/L$, where $L$ is the lattice constant of the single-layer PhC. **e,** Amplitude and phase of theoretically calculated Bessel-type modes for $q_d = 0$ with $n = 0, 1,$ and $2$. **f,** Comparison between simulated and theoretically calculated

mode amplitudes along the Γ-M direction in reciprocal space. The blue solid curve denotes the simulated result, and the orange dashed curve indicates the theoretical prediction. **g, h,** Schematic of vortex beams with distinct OAM topological charges radiated from the Bessel-type modes with $q_d = 0$ (**g**) and $q_d = 1$ (**h**). **i,** Schematic illustration of the mechanism enabling OAM tunability at a fixed wavelength. Curves represent the mode intensity spectra of Bessel-type modes under different configurations of interlayer separation and twist angle. Spectra are vertically offset by 1.1 a.u. for clarity.

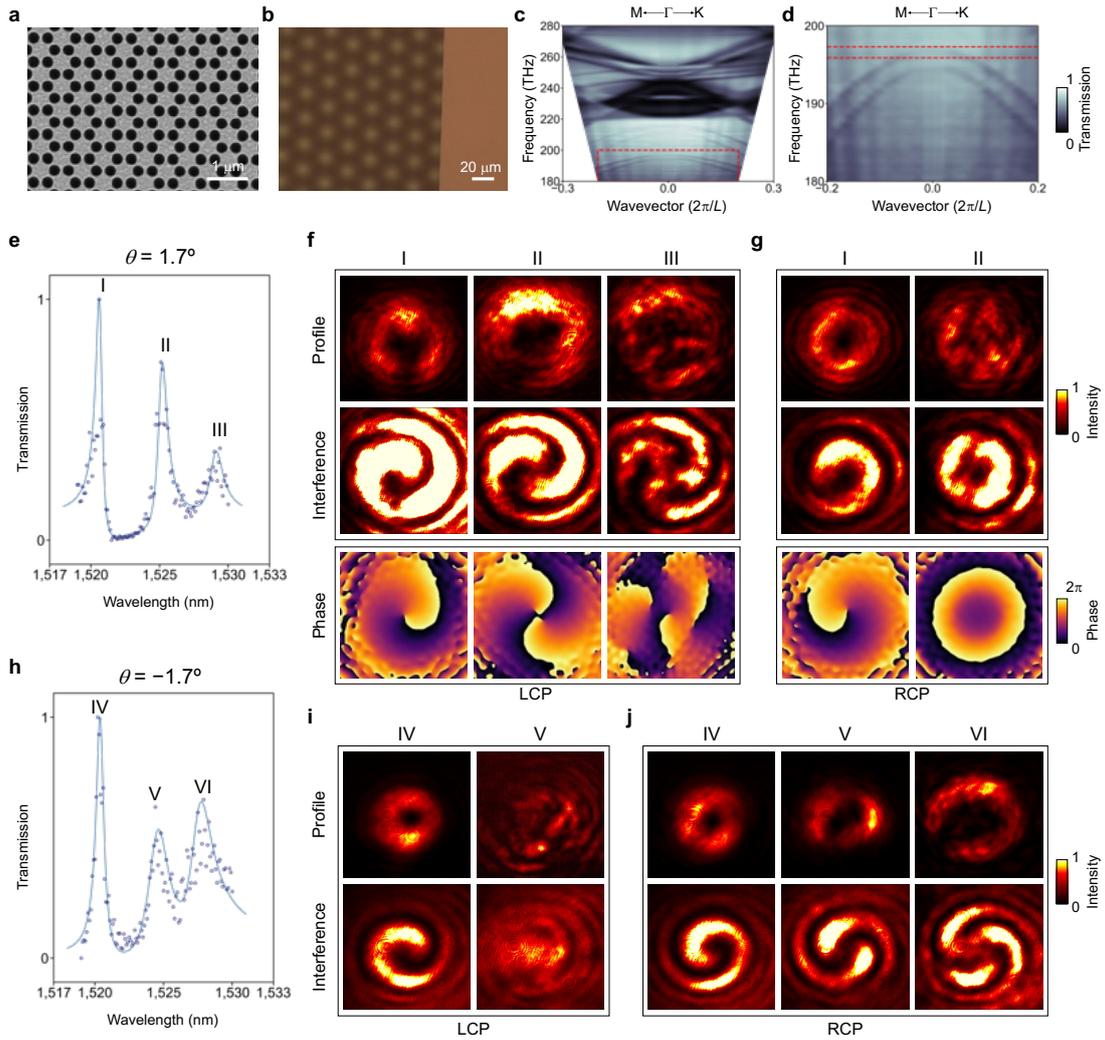

**Fig. 3 | TBMPC sample fabrication and multi-OAM vortex beam generation. a,** Scanning electron microscope (SEM) image of the single-layer PhC slab. **b,** 20× microscope image of the TBMPC, showing a visible moiré pattern. **c,** Experimental band diagram of the TBMPC sample measured via angle-resolved transmittance spectra. The white regions on both sides are beyond the measurement range due to the numerical aperture limitation. The red dashed box indicates the magnified region shown in **d**. **d,** Magnified view of the angle-resolved transmittance spectra near the operating wavelength. The red dashed lines denote the scanned wavelength range from 1,519 nm to 1,530 nm. **e,** Experimentally measured cross-

polarization transmission spectrum of the TBMPC with $\theta = 1.7°$. Dark blue circles: measurement, light blue curve: fitting. **f,** Measured beam profiles (top), measured interference patterns (middle), and simulated phase distributions (bottom) of the left-circularly polarized (LCP) component of the radiation corresponding to peaks I, II, and III in **e**, respectively. **g,** Measured beam profiles (top), measured interference patterns (middle), and simulated phase distributions (bottom) of the right-circularly polarized (RCP) component of the radiation corresponding to peaks I and II in **e**, respectively. **h,** Experimentally measured cross-polarization transmission spectrum of the TBMPC with $\theta = -1.7°$. Dark blue circles: measurement, light blue curve: fitting. **i,** Measured beam profiles (top) and interference patterns (bottom) of the LCP component of the radiation corresponding to peaks IV and V in **h**, respectively. **j,** Measured beam profiles (top) and interference patterns (bottom) of the RCP component of the radiation corresponding to peaks IV, V, and VI in **h**, respectively.

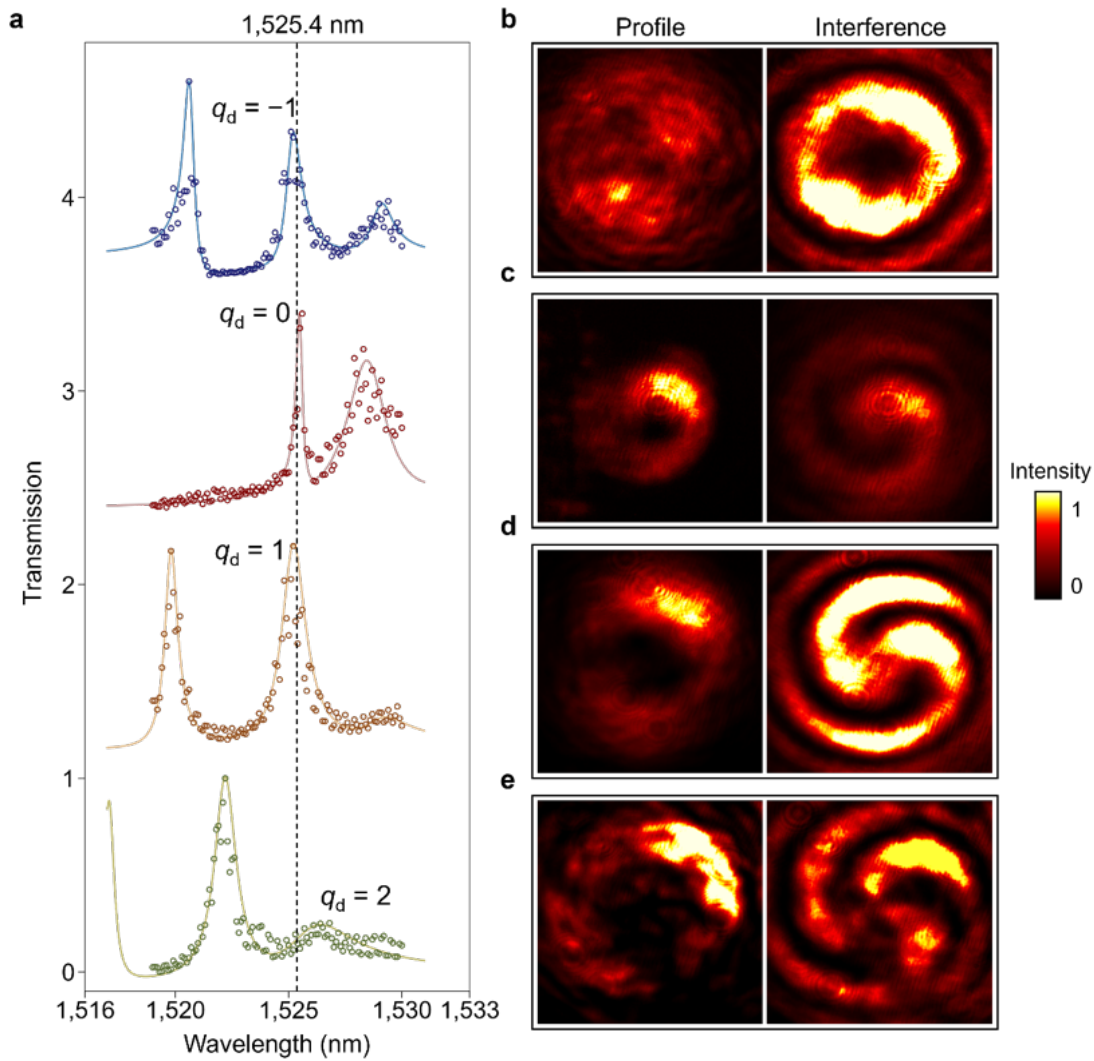

**Fig. 4 | OAM-tunable vortex beam controlled by twist angle and interlayer separation.**

**a,** Experimentally measured cross-polarization transmission spectra of the TBMPC under four distinct configurations of twist angle and interlayer separation. The four spectra (top to bottom) correspond to the configurations: b: ($\theta = 1.7°, t = 280$ nm), c: ($\theta = -1.8°, t = 340$ nm), d: ($\theta = -1.8°, t = 280$ nm), and e: ($\theta = -1.8°, t = 240$ nm) respectively. Circles: experimental data, lines: fitting. The lack of experimental data below 1,519 nm is due to the limited wavelength range of the tunable infrared laser. Spectra are vertically offset by 1.2 a.u. for clarity. **b-e,** Measured beam profiles (left) and interference patterns (right) of the RCP component at $\lambda = 1{,}525.4$ nm for configurations b, c, d, and e, respectively.